\documentstyle{mn}

\newif\ifAMStwofonts

\ifoldfss
  \ifCUPmtlplainloaded \else
    \NewTextAlphabet{textbfit} {cmbxti10} {}
    \NewTextAlphabet{textbfss} {cmssbx10} {}
    \NewMathAlphabet{mathbfit} {cmbxti10} {} 
    \NewMathAlphabet{mathbfss} {cmssbx10} {} 
  \fi
  \ifAMStwofonts
    \ifCUPmtlplainloaded \else
      \NewSymbolFont{upmath} {eurm10}
      \NewSymbolFont{AMSa} {msam10}
      \NewMathSymbol{\upi}     {0}{upmath}{19}
      \NewMathSymbol{\umu}     {0}{upmath}{16}
      \NewMathSymbol{\upartial}{0}{upmath}{40}
      \NewMathSymbol{\leqslant}{3}{AMSa}{36}
      \NewMathSymbol{\geqslant}{3}{AMSa}{3E}

       \let\ge=\geqslant
    \fi
  \fi
\fi 

\ifnfssone
  \newmathalphabet{\mathit}
  \addtoversion{normal}{\mathit}{cmr}{m}{it}
  \addtoversion{bold}{\mathit}{cmr}{bx}{it}
  \newmathalphabet{\mathbfit} 
  \addtoversion{normal}{\mathbfit}{cmr}{bx}{it}
  \addtoversion{bold}{\mathbfit}{cmr}{bx}{it}
  \newmathalphabet{\mathbfss} 
  \addtoversion{normal}{\mathbfss}{cmss}{bx}{n}
  \addtoversion{bold}{\mathbfss}{cmss}{bx}{n}
  \ifAMStwofonts
    \ifCUPmtlplainloaded \else
      \UseAMStwoboldmath
      \makeatletter
      \new@mathgroup\upmath@group
      \define@mathgroup\mv@normal\upmath@group{eur}{m}{n}
      \define@mathgroup\mv@bold\upmath@group{eur}{b}{n}
      \edef\UPM{\hexnumber\upmath@group}
      \new@mathgroup\amsa@group
      \define@mathgroup\mv@normal\amsa@group{msa}{m}{n}
      \define@mathgroup\mv@bold\amsa@group{msa}{m}{n}
      \edef\AMSa{\hexnumber\amsa@group}
      \makeatother
      \mathchardef\upi="0\UPM19
      \mathchardef\umu="0\UPM16
      \mathchardef\upartial="0\UPM40
      \mathchardef\leqslant="3\AMSa36
      \mathchardef\geqslant="3\AMSa3E

       \let\ge=\geqslant
    \fi
  \fi
\fi 

\ifnfsstwo
  \DeclareMathAlphabet{\mathbfit}{OT1}{cmr}{bx}{it}
  \SetMathAlphabet\mathbfit{bold}{OT1}{cmr}{bx}{it}
  \DeclareMathAlphabet{\mathbfss}{OT1}{cmss}{bx}{n}
  \SetMathAlphabet\mathbfss{bold}{OT1}{cmss}{bx}{n}
  \ifAMStwofonts
    \ifCUPmtlplainloaded \else
      \DeclareSymbolFont{UPM}{U}{eur}{m}{n}
      \SetSymbolFont{UPM}{bold}{U}{eur}{b}{n}
      \DeclareSymbolFont{AMSa}{U}{msa}{m}{n}
      \DeclareMathSymbol{\upi}{0}{UPM}{"19}
      \DeclareMathSymbol{\umu}{0}{UPM}{"16}
      \DeclareMathSymbol{\upartial}{0}{UPM}{"40}
      \DeclareMathSymbol{\leqslant}{3}{AMSa}{"36}
      \DeclareMathSymbol{\geqslant}{3}{AMSa}{"3E}

       \let\ge=\geqslant
    \fi
  \fi
\fi 

\ifCUPmtlplainloaded \else
  \ifAMStwofonts \else 
    \def\upi{\pi}
    \def\umu{\mu}
    \def\upartial{\partial}
  \fi
\fi

\title{1.65 $\rm \mu m$ (H-band) surface photometry of galaxies. VIII: \\
the near-IR $\rm \kappa$-space at z=0}
\author[D. Pierini et al.]
       {D. Pierini$^{~1}$, G. Gavazzi$^{~2}$, P. Franzetti$^{~2}$,
M. Scodeggio$^{~3}$ and A. Boselli$^{~4}$ \\
        $^1$Dept. of Physics and Astronomy, The University of Toledo,
2801 W. Bancroft, 43606, Toledo, Ohio \\
        $^2$Universit\'a degli Studi di Milano - Bicocca,
P.zza dell'Ateneo Nuovo 1, I-20126, Milano, Italy \\
	$^3$IFCTR/CNR, Via Bassini 15, I-20133, Milano, Italy \\
        $^4$Laboratoire d'Astrophysique de Marseille, BP8 Traverse du Syphon,
F-13376, Marseille, France}

\date{Accepted ...
      Received ...;
      in original ...}

\pagerange{\pageref{firstpage}--\pageref{lastpage}}
\pubyear{...}

\begin{document}

\maketitle

\label{firstpage}

\begin{abstract}
We present the distribution of a statistical sample of nearby galaxies
in the $\kappa$-space ($\kappa_1 \propto log~M$, $\kappa_2 \propto log~I_e$,
$\kappa_3 \propto log~M/L$). 
Our study is based on near-IR (H-band: $\rm \lambda = 1.65~\mu m$)
observations, for the first time comprising early- and late-type systems. 
Our data confirm that the mean effective dynamical mass-to-light ratio $M/L$
of the E+S0+S0a galaxies increases with increasing effective dynamical mass
$M$, as expected from the existence of the Fundamental Plane relation.
Conversely, spiral and Im/BCD galaxies show a broad distribution
in $M/L$ with no detected trend of $M/L$ with $M$, the former galaxies
having $M/L$ values about twice larger than the latter, on average.
For all the late-type galaxies, the $M/L$ increases with decreasing
effective surface intensity $I_e$, consistent with the existence of
the Tully--Fisher relation.
These results are discussed on the basis of the assumptions behind
the construction of the $\kappa$-space and their limitations.
Our study is complementary to a previous investigation in the optical
(B-band: $\rm \lambda = 0.44~\mu m$) and allows us to study
wavelength-dependences of the galaxy distribution in the $\kappa$-space.
As a first result, we find that the galaxy distribution
in the $\kappa_1$--$\kappa_2$ plane reproduces the transition
from bulge-less to bulge-dominated systems in galaxies of increasing
dynamical mass.
Conversely, it appears that the $M/L$ of late-types is higher (lower)
than that of early-types with the same $M$ in the near-IR (optical).
The origins of this behaviour are discussed in terms of dust attenuation
and star formation history.
\end{abstract}


\section{Introduction}

The study of the scaling relations between photometric
and kinematical properties of present-day stellar systems is a powerful tool
for the comprehension of the processes that led to their formation
and evolution (e.g. Mao \& Mo 1998; Mao, Mo \& White 1998; Pahre,
de Carvalho \& Djorgovski 1998; Avila-Reese \& Firmani 2000).
The relation between the total luminosity $L$ and the central velocity
dispersion $\sigma_0$ (Faber \& Jackson 1976) and the relation between
the effective radius $r_e$ (the radius that contains half of the galaxy
total luminosity) and the central surface brightness $\mu_0$ (Kormendy 1977)
represent benchmarks for any theoretical modelling of the early-type galaxies
(de Zeeuw \& Franx 1991 and references therein).
Further analyses pointed out that these galaxies and the bulges
of the spiral galaxies populate a two-dimensional manifold in the parameter
space defined by $r_e$, $\mu_e$ and $\sigma_0$, named the Fundamental Plane
(FP) (Djorgovski \& Davis 1987; Dressler et al. 1987), where $\mu_e$
is the mean effective surface brightness (mean surface brightness
within $r_e$).
The profound implications of this result, its interpretation
and its extension to parent systems at high redshifts are still grounds both
of observational challange and of intellectually exciting debate
(e.g., Scodeggio et al. 1998; Pahre, Djorgovski \& de Carvalho 1999
and references therein).
For the late-type galaxies, the correlation between the total luminosity $L$
and the maximum rotational velocity $V_{Max}$ (Tully \& Fisher 1977),
as well as the distribution in the parameter space defined by exponential
disk scale-length $h$ and central surface brightness $\mu_0$
(Grosb$\rm \o$l 1985; de Jong 1996a; see also Pierini 1997)
are now compelling for any theory of disk-galaxy formation
(e.g. Dalcanton, Spergel \& Summers 1997; Mo, Mao \& White 1998;
Syer, Mao \& Mo 1999).
In order to reproduce the locus of these systems in the $h$--$\mu_0$ plane,
a bivariate distribution in mass and spin parameter of the proto-galaxies
is required, whatever the model of disk-galaxy formation is (cf. Dalcanton,
Spergel \& Summers 1997; Firmani \& Avila-Reese 2000).

The effective and central photometric parameters involved by the previous
scaling relations were determined traditionally from fitting either
a de Vaucouleurs $r^{1/4}$ law (de Vaucouleurs 1948) or an exponential law
(Freeman 1970) to the radial surface brightness profiles.
The first law was used to describe the light distribution along
the radial coordinate of the elliptical galaxies and of the bulges
of the spiral galaxies, after the pioneering analyses of de Vaucouleurs
(1948, 1959).
The second one was found to be a good description of the light profiles
of the disks, also in the case of S0 galaxies (Burstein 1979).

Burstein et al. (1997 -- hereafter referred to as BBFN; see also
Bender, Burstein \& Faber 1992; Burstein et al. 1995) proposed
a three-dimensional parameter system, called $\kappa$-space, in order
to represent and to compare the global relationships of any stellar system
(i.e., from the Galactic globular clusters to the galaxy clusters),
in a self-consistent way.
The axes of the $\kappa$-space, $\kappa_1$, $\kappa_2$ and $\kappa_3$,
are proportional to the logarithms of the galaxy mass and mass-to-light ratio
and of a third quantity, which is basically the surface brightness,
respectively, and do not depend on the fitting law of the surface brightness
profile.
They reflect a set of physical assumptions on the structure and velocity
pattern of the different self-gravitating, equilibrium stellar systems
(see BBFN), which may be subject of criticism on the basis of some
observational results (e.g. Caon, Capaccioli \& D'Onofrio 1993;
Busarello et al. 1997).
BBFN investigated kinematical and structural properties of nearby galaxies
on the basis of a large optical (B-band: $\rm \lambda = 0.44~\mu m$)
data-set.
Further analyses focussed on the distribution in the near-IR $\kappa$-space
either of present-day E+S0 systems (Pahre, Djorgovski \& de Carvalho
1998 -- hereafter referred to as PDdC) or of disks and bulges
of present-day late-type galaxies (Moriondo, Giovanelli \& Haynes 1999).
The adoption of near-IR photometry gives an insight of the scaling properties
of galaxies less biased by differences in age and metallicity
of the stellar populations (de Jong 1996b) and by dustiness
(Cardelli, Clayton \& Mathis 1989 and references therein) than
the picture emerging from optical studies.

Here we present the near-IR (H-band: $\rm \lambda =1.65~\mu m$)
$\kappa$-space of both early- and late-type galaxies as a whole at z=0.
In Sect. 2 we define the $\kappa$-space axes and their relations
with typical galaxy parameters, such as mass, radius, surface brightness
and luminosity, and discuss the physical assumptions behind
the $\kappa$-space and their limitations.
The selection criteria of our sample and the determination of the galaxy
parameters involved by the definition of the $\kappa$-space axes
are described in Sect. 3.
Here we note that this sample comprises several Virgo cluster
galaxies down to $\rm M_B = -15~mag$ (cf. Boselli et al. 1997, 2000a),
though it is not complete to low surface brightness and dwarf galaxies.
We present and discuss the galaxy distribution in the near-IR $\kappa$-space
in Sect. 4.
Discussion and conclusions on the wavelength-dependence of the galaxy
distribution in the $\kappa$-space are given in Sect. 5.

\section{Physical meaning of the k-space parameters:
assumptions and their limitations}

In origin (Bender, Burstein \& Faber 1992), the $\kappa$-space
is a representation of the structural properties of the elliptical galaxies
(but also of bulges of S0 and spiral galaxies) which makes use
of a particular orthogonal rotation of the global parameter space
defined by the one-dimensional central velocity dispersion ($\sigma_0$),
effective radius ($r_e$) and mean effective surface intensity ($I_e$).
This orthogonal rotation provides face-on and edge-on views
of the Fundamental Plane, under the assumption that the ellipticals
form a homologous family, i.e., that the two structure parameters
$c_1$ and $c_2$, which define the total luminosity $L_T$
($L_T = c_1 I_e r_e^2$) and the total mass $M_T$ ($M_T = c_2 \sigma_0^2 r_e$)
of the system, are constant.
As a consequence, it implies the following logarithmic relations between
the three $\kappa$-coordinates $\kappa_1$, $\kappa_2$ and $\kappa_3$
and the parameters $\sigma_0$, $r_e$ and $I_e$ of the early-type galaxies:
\begin{equation}
\kappa_1 = (log~\sigma_0^2+log~r_{\mathrm{e}}) / \sqrt 2
\end{equation}
\begin{equation}
\kappa_2 = (log~\sigma_0^2+2 log~I_{\mathrm{e}}-log~r_{\mathrm{e}})/ \sqrt 6
\end{equation}
\begin{equation}
\kappa_3 = (log~\sigma_0^2-log~I_{\mathrm{e}}-log~r_{\mathrm{e}})/ \sqrt 3\,,
\end{equation}
where $r_e$ is in kpc, $I_e$ is in solar units and $\sigma_0$
is in $\rm km~s^{-1}$.

While $c_1 = 2 \pi$, by definition of $I_e$, the assumption of a constant
$c_2$ is more uncertain and, therefore, plays a more fundamental role.
Bender, Burstein \& Faber calculated that $c_2/c_1$ decreases by about
a factor 1.6 from dwarf ellipticals to giant ellipticals
in a roughly monotonic fashion, under the assumption that ellipticals
are described by King (1966) models with an isotropic velocity dispersion.
Of course real ellipticals may have more complex internal kinematics.
The debate on homology is still waiting for definitive answers,
but from theoretical (Capelato, de Carvalho \& Carlberg 1995, 1997)
and observational (Graham \& Colless 1997; Busarello et al. 1997) studies
it seems to emerge that kinematical deviations from a homologous family
may produce more important effects than structural deviations
(see also Pahre, de Carvalho \& Djorgovski 1998).

Successively, BBFN proposed the $\kappa$-space as a tool of representation 
of the structural properties of all stellar systems.
In particular, for spiral and irregular galaxies, this required the adoption
of a transformation from their characteristic velocity at $r_e$
to their central velocity dispersion $\sigma_0$, their $r_e$ and $I_e$
being defined as for the ellipticals.
Under the assumption that spiral and irregular galaxies have
exponential Freeman (1970) disks and central surface brightnesses within
1 $\rm mag~arcsec^{-2}$ of 21.65 B-$\rm mag~arcsec^{-2}$,
the characteristic velocity at $r_e$ is reasonably assumed to be equal to
the maximum rotational velocity $V_{Max}$, on the basis of the observational
results of Rubin et al. (1985).
The transformation from $V_{Max}$ to $\sigma_0$ adopted by BBFN
(i.e., $V_{Max}/\sigma_0 = \sqrt 2 $) descends from the further assumption
that disks are embedded in isotropic isothermal halos
(Binney \& Tremaine 1987, Equ. 4--55), and is consistent with observed values
of this ratio (Whitmore \& Kirshner 1981).
With this transformation, the $\kappa$-coordinates of spiral
and irregular galaxies are given by:
\begin{equation}
\kappa_1 = (log~V_{\mathrm{Max}}^2+log~r_{\mathrm{e}}) / \sqrt 2 - 0.21
\end{equation}
\begin{equation}
\kappa_2 = (log~V_{\mathrm{Max}}^2+2 log~I_{\mathrm{e}}-log~r_{\mathrm{e}}) / \sqrt 6 - 0.12
\end{equation}
\begin{equation}
\kappa_3 = (log~V_{\mathrm{Max}}^2-log~I_{\mathrm{e}}-log~r_{\mathrm{e}}) / \sqrt 3 - 0.17\,,
\end{equation}
where $V_{Max}$ is expressed in $\rm km~s^{-1}$.

According to the previous assumptions, the dynamical mass and the dynamical
mass-to-light ratio of any early- and late-type galaxy within $r_e$
may be expressed in terms of the definitions 1--6 of the $\kappa$-coordinates.
The effective dynamical mass (i.e., the mass within $r_e$) is defined by
$M_e= 4.65 \times 10^5 \sigma_0^2~r_e~\rm M_{\odot}$, where
the standard Keplerian formula $M_e=r_e~V^2(r_e)/G$ is adopted in order
to define the mass of a disk, with $V(r_e)$ set equal to $V_{Max}$
and transformed to $\sigma_0$.
As a consequence, the effective dynamical mass (in solar units) is given by:
\begin{equation}
log~M_{\mathrm{e}}(\mathrm{M}_{\odot}) = \sqrt 2 \kappa_1 + 5.67\,.
\end{equation}
From the definition of effective luminosity (i.e., the luminosity within
$r_e$) $L_e = \pi \times 10^6 I_e~r_e^2$,
\begin{equation}
log~L_{\mathrm{e}}(\mathrm{L}_{\odot}) = \sqrt 2 \kappa_1 - \sqrt 3 \kappa_3 + 6.50\,,
\end{equation}
so that the effective dynamical mass-to-light ratio is expressed as:
\begin{equation}
log~M_{\mathrm{e}}/L_{\mathrm{e}}~(\odot) = \sqrt 3 \kappa_3 - 0.83\,.
\end{equation}

In elliptical (and lenticular) galaxies, neglect of rotational support
and flattening may lead to the underestimate of $M$, as derived from
Equ. 7.
The fractional contribution of ordered motion (rotation) to the total kinetic
energy is not negligible in elliptical galaxies of intermediate mass (BBFN).
In recent years (Simien \& Prugniel 2000 and references therein),
it has been found that, in elliptical galaxies, the ratio of
rotational velocity and velocity dispersion has a broad range,
with an average value of the order of 40\%, roughly.
For an isotropic oblate rotator, this figure corresponds to
a fractional change in kinetic energy due to rotation of about 5\%,
which is associated with an ellipticity less than 0.2
(Binney \& Tremaine 1987).
According to the computational results tabulated by the latter authors,
the correction factor for the mass of a rotating oblate galaxy
with the previous characteristic is less than 1.12, i.e., 0.05 dex.
As a reference, for rotating ellipticals of maximum ellipticity $\sim$ 0.4,
the theoretical mass correction is 1.35, i.e., 0.13 dex.

The assumptions made in the case of spiral and irregular galaxies
affect the $\kappa$-coordinates of these stellar systems as well.
One reason of concern is that these objects are far from being
described by a Freeman exponential profile in a homogeneous way,
and not only when a bulge is present.
In fact, in the near-IR, the central surface brightness of the disk-component
of spiral and irregular galaxies falls within a broad range of
$\sim$ 4 $\rm mag~arcsec^{-2}$ (Grosb$\rm \o$l 1985; de Jong 1996a;
Pierini 1997).
Moreover, the effective radius of a spiral galaxy depends on
the bulge-to-disk ratio in a complex way, this ratio depending on
the formation mechanisms of bulge and disk, and on the spectral energy
distribution of each of these two components.
On a statistical basis, Gavazzi et al. (2000b  -- hereafter referred to
as Paper V) and Scodeggio et al. (2001  -- hereafter referred to as Paper IX)
have found that the central near-IR--light concentration increases with
near-IR luminosity in spiral and irregular galaxies.
Consistently,  more luminous systems tend to have a bulge.
This means that, for these stellar systems, the effective radius corresponds
to a lower fraction of the isophotal radius with increasing luminosity.
The radius where the (supposed) universal shape of the rotation curve
of spiral galaxies reaches its maximum velocity changes with luminosity
as well (Persic, Salucci \& Stel 1996), this radius becoming much shorter
than the isophotal optical radius with increasing luminosity.
On one hand, these results support the assumption that
$V(r_e) = V_{Max}$ in spiral galaxies with a bulge.
Therefore, Equ. 7 and 8 may be applied to spiral galaxies with a bulge,
the more luminous ones, on average, the principal unknown being
whether the transformation from $V_{Max}$ to $\sigma_0$ changes
for these systems, i.e., more generally, with luminosity.
On the other hand, the same assumption may be made
for spiral and irregular galaxies of lower luminosities since
the rotation pattern of these objects becomes shallow at radii
$\ge r_e$ (Persic, Salucci \& Stel 1996), with the caveat that,
by imposing $V(r_e) = V_{Max}$, $V(r_e)$ may be significantly overestimated
in the least luminous objects.

\section{Data}

\subsection{Sample and data analysis}

In the last 7 years, Gavazzi and collaborators have carried out an extensive
imaging survey of 1157 early- and late-type galaxies of the Local Universe,
in the near-IR bands H and/or $\rm K^{\prime}$ ($\rm \lambda = 2.10~\mu m$).
The sample galaxies were optically selected in regions of the A262,
Cancer and Virgo clusters, and of the Coma Supercluster (cf. Gavazzi,
Pierini \& Boselli 1996 - hereafter referred to as GPB).
We determine galaxy distances either from individual redshifts
using $\rm H_0 = 75~km~s^{-1} Mpc^{-1}$ (isolated galaxies
of the Coma Supercluster and Cancer cluster members) or via
the cluster assignment method, i.e. from the average redshift
of the structure to which they belong (cf. GPB).
According to the different estimated distances and adopted limiting
magnitudes, only for the Virgo cluster subsample (Boselli et al. 1997,
2000a) the observations reach the bright end of the luminosity function
of dwarf systems.
We identify dwarf systems as those which satisfy the commonly accepted
distance-luminosity definition of Tammann (1980).

The entire survey is described in details elsewhere (Gavazzi et al. 1996b,c;
Boselli et al. 1997; Scodeggio et al. 1998; Gavazzi et al. 2000a;
Boselli et al. 2000a -- hereafter referred to as Papers I and II, B97, S98,
Papers III and IV) and we refer the reader to these papers for details
concerning observations, data reduction, photometric calibration
and image reduction procedures of these 1157 galaxies.
For the present study it is important to say that azimuthally averaged
surface brightness profiles were derived and succesively fitted using
either a de Vaucouleurs $r^{1/4}$ law, an exponential-law, a bulge$+$disk
model or an exponential/de Vaucouleurs truncated model.
The total magnitude of each galaxy was obtained by adding to the flux
measured within the outermost significant isophote the flux extrapolated
to infinity along the model that fitted the outer parts of the galaxy.
The typical internal error in the determination of $\rm H_T$ is $\pm$0.15 mag.
The effective radius $r_e$ and the mean surface brightness within $r_e$
(i.e., $\mu_e$) of each galaxy were computed in two ways (Paper V).
The ``fitted'' values of the individual components were derived from
the individual fitted profiles, extrapolated to zero and to infinity.
By contrast, the ``empirical'' values of the effective radius
and surface brightness of the system were obtained locating
the half-light point along the observed light profile, where the total amount
of light is given by the total magnitude described above, and corrected for
seeing according to Saglia, Bender \& Dressler (1993).
The ``empirical'' values of $r_e$ and $\mu_e$ are used here.
Typical internal errors in the determination of $log~r_e$
and $\mu_e$ are $\pm$0.05 and $\pm$0.16 $\rm mag~arcsec^{-1}$, respectively.
Empirical values of the radii that enclose 75\% and 25\% of the total light
were also determined, and their ratio defines the concentration index
$c_{31}$ (de Vaucouleurs 1977).
Since part of the late-type galaxies in the Virgo cluster were observed
in the $\rm K^{\prime}$-band, their surface brightness profiles
were corrected for the mean H-$\rm K^{\prime}$ color term
found by B97 (0.26 mag) before the determination of $r_e$ and $\mu_e$.

For the late-type galaxies, accurately determined (double or single horned
profile with high signal-to-noise ratio) line-widths (a mean of
the full-widths at the 20\% and 50\% levels of the maximum intensity)
are taken from homogeneous low spatial resolution HI
(i.e., $\rm \lambda = 21~cm$) spectroscopic data-sets available
in the literature (Scodeggio \& Gavazzi 1993 and references therein).
In absence of 21 cm measurements or for HI-deficient spirals
(Haynes \& Giovanelli 1984), maximum rotational velocities are derived
from optical (i.e., $\rm H \alpha$ line, $\rm \lambda = 6563~\AA$)
major-axis rotation curves either derived by us (Gavazzi et al. 1999)
or available in the literature.
Reassuringly, no evidences of systematic differences in the measured
line-widths either from HI or from $\rm H \alpha$ spectroscopic data
are found for the subsample of non HI-deficient objects with
both measurements (Gavazzi et al. 1999).
The values of $V_{Max}$ are derived from the line-widths, corrected for
turbulent/z-motions along the line-of-sight ($\rm -12.6~km~s^{-1}$)
according to Richter \& Huchtmeier (1984) and deprojected according to
the Hubble type-dependent Holmberg formula given by Haynes \& Giovanelli
(1984).
For the early-type galaxies, the values of the velocity dispersion 
are taken from the literature (S98 and references therein)
and homogenized, if necessary, according to the latter authors.

In the present analysis, we exclude the dE and dS0 galaxies observed by
Gavazzi et al. (2001) since it is still matter of debate whether
their stellar disk-component is rotationally flattened or not
(e.g. Rix, Carollo \& Freeman 1999; Geha, Guhathakurta \& van der Marel 2001).
By contrast, we select 197 early-type galaxies and 210 late-type galaxies,
out of the surveyed sample of 1157 galaxies, according to the following
criteria: \newline
1) available homogeneous reliable measurements either of the central
velocity dispersion or of the maximum rotational velocity,
for early- and late-type galaxies, respectively; \newline
2) (only for the late-type systems) galaxy inclination between $\rm 30^o$
and $\rm 85^o$, in order to limit any deprojection bias on $V_{Max}$
and a non-negligible inclination correction of the photometric properties
(if any), respectively; \newline
3) goodness of the mono-dimensional (i.e. radial) surface brightness profile
fit (Paper V), assured by the rejection threshold $\rm \chi^2 = 3.5$.
Though $r_e$ and $\mu_e$ do not come from the fitting of the radial
surface brightness profile, we adopt this criterion in order to select
a sample of ``well-behaved'' galaxies, i.e. of objects with photometric
properties not highly affected either by prominent structural features
different from bulge and disk or by peculiarities.

Requirement 1 limits the statistics of our sample by and large, so that
we do not adopt any cut either in $V_{Max}$ or in $\sigma_0$, though
literature sources suggest that values of $V_{Max} \rm < 100~km~s^{-1}$
and of $\sigma_0 \rm < 100~km~s^{-1}$ are less reliable (but see PDdC).
The availability of reliable measurements of $V_{Max}$ limits the statistics
of galaxies later than Sc but not that of S0a--Sab galaxies.

No completeness is claimed for the resultant sample of 407 galaxies,
which is, therefore, exposed to the cluster population
incompleteness bias (Teerikorpi 1987, 1990).
Since the goal of the present paper is not to determine statistically
accurate relations between the parameters of different scaling relations
but to discuss the relative distribution of each morphological class
of galaxies within the $\kappa$-space, the effects of such a bias
are less worrisome.

For the sample here selected, Table 1 reproduces the relevant information
as follows: \newline
1) the galaxy denomination, from either the ``Catalogue of Galaxies
and Clusters of Galaxies'' of Zwicky et al. (1961--1968 -- CGCG)
or from the ``Virgo Cluster Catalogue'' of Binggeli, Sandage \& Tammann
(1985 -- VCC); \newline
2) the Hubble type ($0 =$ E -- E/S0, $1 =$ S0, $2 =$ S0a -- S0/Sa,
$3 =$ Sa, $4 =$ Sab, $5 =$ Sb, $6 =$ Sbc, $7 =$ Sc (dSc), $8 =$ Scd,
$9 =$ Sd, $10 =$ Sdm -- Sd/Sm, $11 =$ Sm, $12 =$ Im (Im/S), $13 =$ generic
irregulars of Coma and peculiar galaxies, $14 =$ S/BCD (dS/BCD, dS0/BCD,
Sd/BCD), $15 =$ Sm/BCD, $16 =$ Im/BCD); \newline
3) the major axis; \newline
4) the minor axis; \newline
5) the characteristic velocity (i.e., either the central velocity dispersion
$\sigma_0$ or the maximum rotational velocity $V_{Max}$ according
to morphology); \newline
6) the effective radius $r_e$; \newline
7) the effective surface brightness $\mu_e$ (the effective surface intensity
in H-band solar units is given by $I_e = 10^{-0.4 (\mu_e - 24.91)}$);
\newline
8) $\kappa_1$; \newline
9) $\kappa_2$; \newline
10) $\kappa_3$.
%

The values of $r_e$ and $\mu_e$ listed in Tab. 1 are corrected for extinction
as described in the next section.

\subsection{Extinction corrections}

Dust extinction affects the observed global metric and photometric properties
of potentially any galaxy somehow, since dust production is a result of
star formation.
The magnitude of these effects changes with the pass-band of the observations
and depends on the physical and chemical properties of the dust grains,
the topology of the dust, the distribution of the dust with respect
to the different stellar populations and the attenuation optical depth
along the line-of-sight in a non trivial way.
A careful correction of the photometric properties for dust attenuation,
in particular, requires modeling of the stellar populations
and of the radiative transfer within each galaxy, on one hand,
and a multi-wavelength data-set for any individual galaxy, on the other.
This task is prohibitive for a large sample like ours.
Therefore, we have to make a set of assumptions and adopt a set
of corrections based on statistics.
The relatively low exctinction in the H-band (Cardelli, Clayton \& Mathis
1989 and references therein) gives some insurance against large systematic
errors, though the magnitude of the error may vary on an individual basis.

Elliptical and lenticular galaxies may contain dust, either in the form
of lanes and patches or of a diffuse component of the interstellar medium
(Bertola \& Galletta 1978; Hawarden et al. 1981; Jura 1986).
The wavelength-dependent effects of dust absorption and scattering
on their photometric properties have been studied by Witt, Thronson
\& Capuano (1992) under the assumption of a diffuse dust component.
As shown by these authors, attenuation optical depths of 0.1--1 in the V-band
($\rm \lambda =0.55~\mu m$) may explain some of the photometric properties
of these stellar sytems, e.g. their color gradients.
Starlight attenuation is primarily the result of absorption in the near-IR,
where the absorption optical depths are lower than in the optical
(Witt, Thronson \& Capuano).
Dust is found capable to affect the projection
of the kinematical quantities of these stellar systems as well
(Baes \& Dejonghe 2000; Baes, Dejonghe \& De Rijcke 2000).
Nonetheless, under the assumption that only the diffuse dust component
is present in elliptical and lenticular systems, we may reasonably assume
that these stellar systems are basically dust-free in the H-band,
in agreement with Baes \& Dejonghe.

This assumption is questionable for spiral and irregular galaxies.
The debate on the opacity of the disk-component of spiral and irregular
galaxies is still open and its discussion is beyond our scope
(see Disney et al. 1989 for a review).
If dust is distributed only in the disk, disks are more absorbed
at high inclinations, because of the larger optical depth of the dust
in the plane of the galaxy.
By contrast, bulges are more attenuated at low inclinations, as in this case
basically only the foreground light of the system is observed, whereas
at high inclinations, almost the entire bulge comes into view,
if the dust thickness is not substantial with respect to the bulge
effective radius (Ferrara et al. 1999).
As a consequence, in bulge+disk systems, the dust effects on
$r_e$ and $\mu_e$ will depend not only on the opacity along the line-of-sight
but also on the bulge-to-disk ratio at a given pass-band.
To make things more complex, bulges may be dusty too (Peletier et al. 1999).
For bulge+disk systems, recipes for the correction of $r_e$ and $\mu_e$
for dust attenuation, motivated either by radiative transfer models
or by observational studies, still lack.
By contrast, dust attenuation in exponential disks has driven more attention
both from theory (e.g. Bianchi et al. 2000) and observations
(e.g. Moriondo, Giovanelli \& Haynes 1999).
For the purposes of our study, we make the following assumptions:
i) dust is distributed only in the disk-component of spiral and irregular
galaxies; ii) the inclination corrections for $r_e$ and $\mu_e$
are the same as for the exponential disk scale-length ($h$)
and central surface brightness ($\mu_0$), respectively, whatever
the late Hubble type of the galaxy is.
Given the previous discussion, we aknowledge that the second assumption
over-simplifies the problem.
However, at least this assumption does not lead to an inconsistency when
we compare the effective luminosities obtained through Equ. 10, after
applying the extinction correction of Moriondo, Giovanelli \& Haynes (1999)
to $r_e$ and $\mu_e$, and the total luminosities obtained from our data
analysis and corrected for internal extinction (Galactic extinction
is negligible in the H-band) according to the recipes of Gavazzi \& Boselli
(1996).
The corrections for inclination applied to $r_e$ and $\mu_e$ are:
\begin{equation}
r_{\mathrm{e,c}} = r_{\mathrm{e,o}} / (1+0.41 (\pm 0.17) log~a/b)
\end{equation}
\begin{equation}
\mu_{\mathrm{e,c}} = \mu_{\mathrm{e,o}} + 1.5 (\pm 0.35) log~a/b\,,
\end{equation}
where $r_{e,o}$, $r_{e,c}$, $\mu_{e,o}$ and $\mu_{e,c}$ are the observed
and corrected values of the effective radius (in arcsec) and of
the mean effective surface brightness (in $\rm mag~arcsec^{-2}$),
respectively, and $a/b$ is the major-to-minor axis ratio.
Total magnitudes are corrected for internal extinction as follows
(Gavazzi \& Boselli 1996):
\begin{equation}
m_{\mathrm{T,c}} = m_{\mathrm{T,o}} -2.5 \times 0.17 log~a/b\,,
\end{equation}
where $m_{T,o}$ and $m_{T,c}$ are the observed and corrected values,
whatever the Hubble-type is.
In fact, Gavazzi \& Boselli did not find any morphological dependence
of the corrections of total magnitudes for internal extinction
in the near-IR pass-bands, as opposed to the optical ones.

\section{The H-band k-space}

Fig. 1 shows the distribution of the sample galaxies in a three-dimensional
fold-out of the near-IR $\kappa$-space.
Here we group the sample galaxies in ellipticals (E) plus lenticulars (S0),
S0a galaxies, and giant spirals (Sa--Sc) plus late-spirals/Im/BCDs
(Scd--Im/BCD) and generic irregular/peculiar galaxies.
E+S0, S0a, Sa--Im/BCD galaxies plus generic irregular/peculiar galaxies
are represented with filled circles, asterisks and empty circles,
respectively, in Fig. 1.
We assume characteristic cumulative uncertainties
$\delta \kappa_1 = \delta \kappa_2 = \delta \kappa_3 = 0.1$,
consistent with typical uncertainties in the observables (cf. Sect. 3).
The magnitudes of these 1 $\rm \sigma$ errors are reproduced in each panel
of Fig. 1.
We also reproduce the increase in all the three $\kappa$-space coordinates
due to the increase of the dynamical mass when the potential contribution
of rotational velocity to the total kinetic energy of the ellipticals
is taken into account (cf. Sect. 2).

Here we need to say that the impact of the extinction corrections
on the distribution of the late-type galaxies in the near-IR $\kappa$-space
is small but the fact that the average effective mass-to-light ratio
of these galaxies increases by 70\% when extinction corrections are applied.

In the next subsections we analyse the galaxy distribution
in each projection of the $\kappa$-space as a function of morphology.
Therefore, in Fig. 2, we reproduce the three-dimensional fold-out
of the near-IR $\kappa$-space for these four groups of Hubble types:
\begin{itemize}
\item
E, S0 and S0a, represented by filled circles, empty squares and asterisks,
respectively (panel a);
\item
Sa+Sab and Sb, represented by broad crosses and empty hexagons, respectively
(panel b);
\item
Sbc and Sc, represented by filled triangles and empty circles, respectively
(panel c);
\item
Scd, Sd and Im/BCD plus generic irregular/peculiar galaxies,
represented by empty pentagons, stars and crosses, respectively (panel d).
\end{itemize}
We adopt this classification from BBFN (except for the generic
irregular/peculiar galaxies).

In each panel of Fig. 2, we reproduce the Fundamental Plane relation
in the $\kappa_1$--$\kappa_3$ plane (see Sect. 4.1.1), the distribution
of the E+S0 galaxies, within $\rm \pm 1~\sigma$ from the mean,
in the $\kappa_2$--$\kappa_3$ plane (see Sect. 4.2.1) and the borderline of
the ``Zone of Exclusion'' in the $\kappa_1$--$\kappa_2$ plane
(see Sect. 4.3.1).

\subsection{Galaxies in the $\kappa_1$--$\kappa_3$ plane}

\subsubsection{Ellipticals and lenticulars}

The distribution of the elliptical and lenticular galaxies
in the $\kappa_1$--$\kappa_3$ projection of the $\kappa$-space confirms
that the dynamical mass-to-near-IR light ratio increases with
the dynamical mass (cf. PDdC) as found in the optical (Bender, Burstein
\& Faber 1992; Burstein et al. 1995; BBFN).
This result was expected from the existence of the Fundamental Plane relation
in its canonical notation (Pahre, de Carvalho \& Djorgovski 1998; S98).
A linear fit to the data which minimizes the dispersion in both axes gives:
\begin{equation}
\kappa_3 = 0.242(\pm 0.022) \kappa_1 -0.373 (\pm 0.076)\,.
\end{equation}
The slope of the H-band FP (in $\kappa$-space notation) that we derive
corresponds to the scaling relations $M_e/L_e \propto M_e^{0.296 \pm 0.027}$
and $M_e/L_e \propto L_e^{0.421 \pm 0.055}$.
In the K-band PDdC obtained $M_e/L_e \propto M_e^{0.147 \pm 0.011}$
and $M_e/L_e \propto L_e^{0.172 \pm 0.013}$.
The origin of the significant discrepancy between these two sets of near-IR
scaling relations is twofold: the incompleteness bias and differences
in data analysis.
First we note that both these analyses rest on uncomplete samples,
so that the derived scaling relations are biased.
The PDdC sample is more than twice in size than ours, containing 251
elliptical and lenticular galaxies, selected primarily from 13 nearby
rich clusters (but also from loose groups and the general field),
whose distribution in redshift velocity brackets the redshift velocities
of Virgo and Coma, the main contributors of objects in our study.
According to the study of the incompleteness bias by Teerikorpi (1987, 1990),
we may allow that, in our sample, the fraction of low-luminosity
(and low-mass) objects is skew towards higher luminosities (and masses)
than in the PDdC sample, so that a steeper relation between mass-to-light
ratio and mass may be expected.
Nonetheless, systematic differences in the data analysis between us and PDdC
probably have a larger impact on the obtained scaling relations.
While the seeing corrections adopted by us and PDdC are consistent
(cf. Pahre 1999), the effective radii and mean surface brightnesses
are derived in two different ways.
We obtain these effective parameters from isophotal, elliptical surface
photometry (cf. Sect. 3.1), while PDdC determined them from circular aperture
photometry, in agreement with previous optical studies.
As discussed by Pahre (1999), the isophotal estimate of
the effective radius is, on average, slightly larger than its circular
aperture estimate, and, conversely, the isophotal estimate of the effective
intensity is lower than the corresponding circular aperture estimate.
This was already proposed by S98 in order to explain the discrepancy
between the H-band and the K-band FP relations, the K-band effective
parameters being obtained from the circular aperture estimates
of Mobasher et al. (1999).
If the comparison of these two data-sets is held as representative
of the systematic effects due to the different data analysis, we infer
that the adoption of circular aperture estimates of the effective parameters
causes a systematic decrease of the $\kappa_1$ and $\kappa_3$ coordinates
(cf. Equ. 1 and 3).
This results in a reduced slope of the FP in the $\kappa$-space notation
and contributes to justify the discrepancy between us and PDdC.

\subsubsection{Spirals and Im/BCDs}

Spiral and Im/BCD galaxies form two distinct classes
in the $\kappa_1$--$\kappa_3$ plane, the latter having a mean effective
mass-to-light ratio (in solar units) larger by a factor of 2.
Taken as a whole, the late-type galaxies have a mean value
of $\kappa_3$ equal to 0.656, i.e., a mean effective mass-to-light ratio
(in solar units) equal to 2.
This figure is 43\% of the value of the mean total mass-to-light ratio
(in solar units) determined by GPB for a sample of 426 late-type galaxies.
This large discrepancy has not a statistical origin.
First, the total mass-to-light ratios of GPB were determined, under the same
assumptions of virial equilibrium and spherical symmetry as in Sect. 2,
at the 25 B-$\rm mag~arcsec^{-2}$ isophotal radius $R_{opt}$,
and not at infinity.
As a consequence, the mass-to-light ratios quoted in GPB are about 10\% less
than the total ones, on average. 
From the definitions of total and effective mass and luminosity, $M_T$
and $L_T$ respectively, it is straightforward to determine that
$M_e/L_e = 2 r_e/R_{opt} \times M_T/L_T$.
Whatever the Hubble type of the galaxy is, on average,
$R_{opt}/r_e = 5.3 \pm 0.08$ (Paper V), so that we expect that
the mean effective mass-to-light ratio is 38\% of the mean total
mass-to-light ratio, on average.
Second, we note that the maximum rotational velocities
were not corrected for turbulent/z-motions by GPB, so that
the estimates of the total mass-to-light ratios further exceed those
of the effective mass-to-light ratio for the low-mass objects.
We conclude that the two estimates are consistent within the assumptions
and the uncertainties of the observations.

What do these two estimates tell us about the radial behaviour
of the mass-to-light ratio?
The mass of a spiral galaxy within a given galactocentric distance
lays with high confidence between the estimates given by a flat disk model
and a spherical one and is between 60 and 100\% of the mass estimate
determined through the assumption of spherical symmetry (Lequeux 1983).
According to this author, the overestimate due to this assumption
is probably much reduced at galactocentric distances as large as
the optical radius, where spherical components dominate the potential.
Hence, the 60\% difference between the average estimates of the effective
and total mass-to-light ratios may suggest that the actual mass-to-light ratio
increases with galactocentric distance.

A linear fit to the data of all the galaxies later than S0a
which minimizes the dispersion in both axes gives:
\begin{equation}
\kappa_3 = -0.076(\pm 0.021) \kappa_1 +0.904 (\pm 0.070)\,.
\end{equation}
Formally this relation corresponds to the scaling relations
$M_e/L_e \propto M_e^{-0.093 \pm 0.026}$
and $M_e/L_e \propto L_e^{-0.103 \pm 0.032}$.
The slight dependences of the effective mass-to-light ratio
on effective mass and luminosity are marginally significant.
We note that the adoption of velocity corrections for turbulent/z-motions
gives us some protection against a systematic overestimate of the mass
of the Scd--Im/BCD galaxies, where such motions are not negligible
with respect to the rotational velocity.
The incompleteness bias which affects our sample may justify even higher
mass-to-light ratios for these galaxies, since we better observe
low-mass objects of higher luminosity.
On the other end, Lequeux (1983) shows that the discrepancy between
the estimates of the mass-to-light ratio given by a flat disk model
and a spherical one depends on its rotation curve.
In a disk where the rotation curve increases linearly until a radial distance
$a$, where it reaches its maximum, and stays flat at larger radii,
this assumption leads to a discrepancy increasing from 40 to 50\%
when $a$ ranges between 0 and the optical radius. 
Since rotation curves seem to peak at larger radial distances
with decreasing mass of the system (Persic, Salucci \& Stel 1996),
it is reasonable to assume that our estimates of the effective mass-to-light
ratio for the Scd--Im/BCD galaxies may be systematically in excess
by an additional 10\% with respect to those of earlier and more massive
late-types.
Furthermore, setting the velocity at the effective radius equal to
the maximum velocity may lead to an overestimate of the dynamical mass
of these systems, as discussed in Sect. 2.
Alternatively, the decrease of the dynamical mass-to-light ratio with mass
may be attributed to increasing extinction, as an effect of
the mass-metallicity relation (Zaritsky, Kennicutt \& Huchra 1994). 
At present, there are no estimates of such an effect, obtained from modelling
both kinematics and radiative transfer and we assume that these effects
are very small in the near-IR.
Hence, we conclude that the validity and interpretation of
the scaling relations involved by Equ. 14 are dubious.

\subsection{Galaxies in the $\kappa_2$--$\kappa_3$ plane}

\subsubsection{Ellipticals and lenticulars}

In this projection of the $\kappa$-space, the distribution of
the E and S0 galaxies is elongated in the direction of the $\kappa_2$ axis,
so that $I_e \propto (M_e/L_e)^{-3}$ for these objects (Fig. 2a).
A similar distribution is found by PDdC and BBFN.
Therefore, we conclude that the higher is the mass of
an elliptical/lenticular galaxy the higher is its mean effective
mass-to-light ratio and the lower is its effective surface intensity,
whatever the pass-band is.
S0a galaxies distribute in a similar way to the early-types.

\subsubsection{Spirals and Im/BCDs}

For the late-type galaxies, the coordinates $\kappa_2$ and $\kappa_3$
are not independent.
As shown by BBFN, the distribution of the spiral and Im/BCD galaxies
in this plane reproduces the generalized Tully--Fisher (TF) relation
$log~L_T = A_C~log~V_{Max} + const.$, where $A_C$ is the color-dependent
exponent in the standard TF power law.
In the near-IR $A_C=4$ (Aaronson et al. 1979) whatever the morphological type
is (e.g. Pierini \& Tuffs 1999) and, as a consequence, spiral
and Im/BCD galaxies distribute in a plane that projects
with minimal scatter onto $\kappa_2$--$\kappa_3$ (BBFN).
For the present sample of late-type galaxies we obtain $A_C=3.851 \pm 0.074$,
consistent with the standard exponent of the near-IR TF relation,
but the scatter in the $\kappa_2$--$\kappa_3$ plane is still large.
We conclude that, whatever the astrophysical origin of the TF relation is
(e.g. Silk 1997; Avila-Reese \& Firmani 2000), its existence implies
a non-linear relation between effective surface intensity
and mass-to-light ratio along the whole Hubble sequence of
the spiral galaxies (Pierini \& Tuffs 1999).
The exponent of this power law depends on the observing wavelength
and is $\sim$ 1.5 in the near-IR.

\subsection{Galaxies in the $\kappa_1$--$\kappa_2$ plane}

\subsubsection{Ellipticals and lenticulars}

The $\kappa_1$--$\kappa_2$ plane may be considered a sort of face-on
projection of the galaxy distribution in the $\kappa$-space (BBFN),
where early- and late-type galaxies distribute in two broad
and almost perpendicular regions.
Elliptical and lenticular galaxies populate the broad region defined by
$\kappa_1 > 2.5$ and  $\kappa_2 > 3$, with $\kappa_1 + \kappa_2 < 9$,
approximately, together with the S0a galaxies.
This result is consistent with the distribution of similar galaxies
in PDdC and BBFN.
In analogy with the latter authors we call ``Zone of Exclusion''
the region with $\kappa_1 + \kappa_2 \ge 9$.

\subsubsection{Spirals and Im/BCDs}

The galaxies with morphological type later than S0a populate a broad region
defined by $-1 < \kappa_2 - \kappa_1 < 2$, with $\kappa_1 + \kappa_2 < 9$.
Individual late-type galaxies distribute inside this strip and further
away from the ``Zone of Exclusion'' according to the progression
in morphological type from Sa to Im/BCD, as found in the optical (BBFN).
Consistently, late-type galaxies with decreasing values of
the light concentration index $c_{31}$ populate regions further away
from the ``Zone of Exclusion''.
By contrast, all our E+S0+S0a galaxies have $c_{31} > 2.82$,
the typical value for a pure exponential disk-system, and do not show
any relation between their location in the $\kappa_1$--$\kappa_2$ plane
and their value of $c_{31}$.

\subsubsection{The ``generalized'' Kormendy relation}

The similar behaviours of early- and late-type galaxies in the optical
and near-IR $\kappa_1$--$\kappa_2$ planes must reproduce a change
in the relation between fundamental structural parameters, independent of
the observational wavelength where these parameters are determined.
As suggested by BBFN, the distribution of giant and dwarf early-type galaxies
in the optical $\kappa_1$--$\kappa_2$ plane is analogous to the division
of these stellar systems found by Kormendy (1988), the behaviour of
the Sa--Sc galaxies resembling that of the giant ellipticals and lenticulars.

Therefore, we plot $\mu_e$ vs. $r_e$ for the pure de Vaucouleurs systems
(filled hexagons) in Fig. 3a and for all the E+S0 galaxies of our sample
(filled circles and empty squares, respectively) in Fig. 3b.
The Kormendy relation (Kormendy 1977, 1988; Burstein 1979) of the E galaxies
of the present sample is reproduced in each panel:
\begin{equation}
\mu_{\mathrm{e}} = 2.65 (\pm 0.07) log~r_{\mathrm{e}}+15.65 (\pm 0.04)\,.
\end{equation}
It is no surprise that the subsample of the pure de Vaucouleurs systems
follows such a relation, since this is a natural consequence
of the $r^{1/4}$ law of their light profiles (Khosroshahi, Wadadekar
\& Kembhavi 2000).
However, all the E+S0 galaxies distribute along the same mean relation
(Fig. 3b), whatever their radial surface brightness profiles are.
Hence we conclude that the Kormendy relation is not an artifact
of the $r^{1/4}$ law but it expresses a unique property of all
these stellar systems.

In addition, we plot $\mu_e$ vs. $r_e$ for the pure exponential
disk-systems plus the truncated-disk systems (empty triangles in Fig. 3c)
and for all the de Vaucouleurs/exponential bulge+exponential disk systems
(filled squares in Fig. 3d).
We find that the distribution of the late-type galaxies as a whole class
in the near-IR $r_e$--$\mu_e$ plane is similar to the distribution
of their disk-components in the near-IR and optical $h$--$\mu_0$ planes
(Grosb$\rm \o$l 1985; de Jong 1996a; Dalcanton, Spergel \& Summers 1997;
Pierini 1997).
In fact:
\begin{itemize}
\item
the mean effective surface intensity of spiral systems as a whole
and the central intensity of their exponential disk-components range
within a broad but upper limited interval for any value of
the effective radius and of the exponential disk scale-length, respectively;
\item
in both distributions the limiting intensity is constant for systems
with characteristic sizes less than $\rm \sim 1~kpc$, but decreases
in systems of larger characteristic sizes.
\end{itemize}

It is interesting to realize that the Kormendy relation of
the ellipticals represents the borderline of the broad region
in the $r_e$--$\mu_e$ plane populated by the late-type galaxies.
In particular, the bulge+disk late-type galaxies populate a strip parallel
and immediately below the locus defined by the Kormendy relation, consistent
with Khosroshahi et al. (2000), while the pure exponential/truncated disks
populate the whole permitted region of the $r_e$--$\mu_e$ plane.

The comparison of Fig. 1 and 3 may suggest that structure changes
with the mass of the stellar system (Paper V and IX), under the assumption
of homologous classes of galaxies in gravitational equilibrium,
and that this equilibrium holds only outside the ``Zone of Exclusion'' (BBFN).

\section{Discussion and conclusions}

The existence of the Hubble sequence of galaxies (Hubble 1926; Sandage 1961)
reflects differences in structure, kinematics and global star-formation
histories of the classified stellar systems (Roberts \& Haynes 1994).
The study of the galaxy distribution in the so-called $\kappa$-space (BBFN)
contributes to the understanding of the galaxy phenomenology,
under the non trivial assumption that early- and late-type galaxies
form two distinct homologous classes in gravitational equilibrium.
Our investigation of the $\kappa$-space of nearby galaxies is based on
near-IR surface photometry and is complementary to the near-IR studies
of PDdC (limited to ellipticals and lenticulars) and Moriondo, Giovanelli
\& Haynes (1999) (limited to bulges and disks of late-type galaxies)
and to the optical study of BBFN, though the photometric parameters
are derived via different procedures in these four studies.

As a first result, we find that galaxies of the same morphological type
distribute in the $\kappa_1$--$\kappa_2$ projection of the $\kappa$-space
in a way analogous to the optical case.
This distribution reproduces the ``generalized'' Kormendy relation (BBFN)
since it finds analogies with the Kormendy relation of elliptical
and lenticular galaxies (Kormendy 1977, 1988; Burstein 1979) on one hand
and, on the other, the distribution of the disk-components of late-type
galaxies in the $h$--$\mu_0$ plane (Grosb$\rm \o$l 1985; de Jong 1996a;
Pierini 1997).
It traces changes in the global structure associated with the dynamical mass
of the galaxy (cf. Paper V and IX), beyond any possible systematic effect
due to differences in the radial distribution of the stellar populations
which dominate the light emission in different pass-bands, whether
these differences are intrinsic or due to differential attenuation by dust
(cf. Witt, Thronson \& Capuano 1992 for the early-type galaxies).
We also confirm that the distribution of the early-type galaxies
in the $\kappa_1$--$\kappa_3$ plane reproduces their Fundamental Plane
relation and that, conversely, the distribution of the late-type galaxies
in the $\kappa_2$--$\kappa_3$ plane reproduces the Tully--Fisher relation
of the latter stellar systems.
This time, the galaxy distribution in the $\kappa_1$--$\kappa_3$
and $\kappa_2$--$\kappa_3$ projections of the $\kappa$-space depends
on the pass-band adopted in order to determine the photometric properties
of the galaxies, as a consequence of the dominant role
of the mass-to-light ratio.

The Fundamental Plane relation (in $\kappa$-space notation) shows that
the dynamical mass-to-light ratio of elliptical and lenticular galaxies
increases with dynamical mass, both in the optical and in the near-IR.
The neglect of dust effects, even for rather modest amounts of dust,
leads to an overestimate of the total dynamical mass-to-light ratio
by a factor of $\sim$ 20\% per optical depth unit (Baes, Dejonghe
\& Rijcke 2000).
Therefore, this relation holds almost unaffected by dust bias
of the photometric properties in the near-IR.
By contrast, the neglect of the rotational velocity on the mass estimate
affects the Fundamental Plane relation, whatever the photometric pass-band is.
We estimate that the systematic underestimate of the dynamical mass
of the ellipticals is about 10\%, on average.
As a consequence, in the near-IR, a giant elliptical galaxy would have
a corrected mean effective mass-to-light ratio of about 1.6 (in solar units),
on average, still lower than the average mean effective mass-to-light ratio
of late-type galaxies.
The correction for the lenticular galaxies is not straightforward,
but we may expect that it ranges between 10 and 30\%, on average,
the latter value applying to rotating ellipticals of maximum ellipticity
$\sim$ 0.4.
As a consequence, in the near-IR, the corrected mean effective mass-to-light
ratios of lenticular galaxies might be intermediate between those of earlier
and later Hubble types.

Spiral and Im/BCD galaxies show very different behaviours
in the optical and near-IR $\kappa_1$--$\kappa_3$ planes.
First of all, they have mean effective mass-to-light ratios higher
than those of elliptical and lenticular galaxies of the same mass
in the near-IR but it is vice versa in the optical (cf. BBFN).
A study of the non-trivial effects of dust attenuation on
the surface photometry of the late-type galaxies is beyond the reach
of this analysis and, therefore, we have assumed statistically-based
inclination corrections of the near-IR photometric parameters
of these stellar systems (Equ. 10 and 11).
In the near-IR, these corrections produce an increase of the mean effective
mass-to-light ratio (Fig. 3a,b), opposite to what expected for early-type
galaxies, so that we are confident that the relative distribution
of early- and late-type galaxies $\kappa_1$--$\kappa_3$ plane is robust.
Second, in the optical, the mass-to-light ratio vs. mass relation
shows a strong dependence on the Hubble type of the spiral
and Im/BCD galaxies.
In the near-IR this dependence is not detected (if any) though there may be
a hint that Im/BCD galaxies have higher mass-to-light ratios than spirals.

If not due to systematic differences either in the statistically-based
corrections for dust effects and inclination adopted by us and BBFN
or in the estimate of the photometric parameters, the different locations
of early- and late-type galaxies in the optical and near-IR
$\kappa_1$--$\kappa_3$ planes may originate from residual dust effects
and/or from differences in the characteristic stellar populations,
both dependent on morphology and mass.
The stellar mass-to-light ratio of a simple stellar population (SSP)
increases with age both in the optical and in the near-IR,
its early-time evolution being very fast until $\sim 5$ Gyrs
and becoming mild afterwards (e.g. Maraston 1998).
In particular, the stellar mass-to-near-IR light ratio reaches a value
within 20\% from the present one ($\sim 1.2$ at 15 Gyrs) after only 3 Gyrs,
while the stellar mass-to-optical light ratio amounts to 80\%
of the present one ($\sim 10$) after 13 Gyrs.
Of course a galaxy is not reproduced by a SSP but by a mix of different SSPs,
weighted by its global star formation history.
In particular, star formation activity leads to dust production,
so that the mass-to-light ratio of a more recently born SSP
will be reduced by dust attenuation both at the stellar photosphere
and in the interstellar medium.
It is commonly accepted that the bulk stellar population of giant elliptical
and lenticular galaxies is older than those of spiral and Im/BCD
galaxies, i.e. that the former have transformed gas into stars much faster
than the latter (e.g. Renzini 1998).
An analogous trend of decreasing ages of the characteristic stellar population
with later Hubble type is suggested for the late-type galaxies
(e.g. Sandage 1986; Kennicutt, Tamblyn \& Congdon 1994;
Gavazzi \& Scodeggio 1996; Boselli et al. 2001).
If these considerations apply, we expect that the early-type galaxies
have higher stellar mass-to-light ratios than later ones, where star formation
is still going on, under the assumptions that galaxies of all morphological
types have the same age and the same initial mass function.
In particular, the stellar mass-to-near-IR light ratios of nearby early-
and late-type galaxies will not differ much if the peak of star formation
activity took place more than 3 Gyrs ago for all of them.
By contrast, the stellar mass-to-optical light ratios of early- and late-type
galaxies may still differ by a maximum factor of $\sim$ 7 (Maraston 1998),
if this peak took place not less than 3 Gyrs ago for all of them.

We observe that the dynamical mass-to-near-IR light ratio of elliptical
and lenticular galaxies is lower than that of spiral of the same
dynamical mass.
From the previous considerations, we conclude that the dynamical-to-stellar
mass ratio of the former galaxies is lower than that of the latter,
if the peak of the star formation activity took place more than 3 Gyrs ago
for all of them.
Extending the same conclusion to Im and BCD galaxies is dangerous,
since the near-IR luminosity may be seriously contaminated by emission
from younger stellar populations.
Since BBFN observe that the dynamical mass-to-optical light ratio
of elliptical and lenticular galaxies is higher than that of spiral
galaxies of the same dynamical mass, this behaviour is consistent
with the previous conclusion if the difference in the stellar mass-to-optical
light ratio, due to the different star formation histories of these two
classes of stellar systems, is larger than the difference in
the dynamical-to-stellar mass ratio.

Finally, we note that the optical and near-IR Fundamental Plane relations
should be analogous if the peak of star formation took place more
than 3 Gyrs ago for all the elliptical and lenticular galaxies,
but for a color term, when the differential effects of diffuse dust
on the photometric parameters are taken into account.
Whether this color term is not or is constant, it depends on the existence
of the color-magnitude relation or not (Scodeggio 2001 and references
therein).
The answer to this question would help understanding whether
the Fundamental Plane relation (in $\kappa$-space notation) is (mainly) due
to the increase either of the dynamical-to-stellar mass ratio or of the
stellar mass-to-light ratio with dynamical mass.

\section*{Acknowledgments}
D.P. wishes to thank David C. Koo and Claudia S. M\"oller
for illuminating discussions held at the G\"ottingen Sternwarte.
\newline
We are thankful to the several night-assistents, undergraduate students
and researchers who contributed to the near-IR survey at the basis
of this work.
\newline
We are gratefult to the anonymous referee, whose stimulating comments
and suggestions led to the improvement of the original manuscript.


\section*{Figure captions}
{\bf Fig. 1:} The three-dimensional fold-out of the $\kappa$--space,
where $\kappa_1 \propto log~M$, $\kappa_2 \propto log~I_e^3 \times M/L$
and $\kappa_3 \propto log~M/L$.
Here, we represent elliptical (E) and S0 galaxies with filled circles,
S0a galaxies with asterisks and Sa -- Im/BCD galaxies with empty circles.
In each panel, we represent characteristic observational errors (crosses)
and the expected increase of the three $\kappa$-coordinates
of E and S0 galaxies when the kinetic energy of these stellar systems
is not negligible.
Late-type galaxies have higher values of $M/L$ than earlier ones
of the same mass, contrary to what found by BBFN in the optical.
\newline
{\bf Fig. 2:} The three-dimensional fold-out of the $\kappa$--space
for different groups of morphological types is reproduced.
In particular, we represent: E, S0 and S0a galaxies with filled circles,
empty squares and asterisks, respectively ({\bf a}); Sa$+$Sab and Sb galaxies
with broad crosses and empty hexagons, respectively ({\bf b}); Sbc and Sc
galaxies with filled triangles and empty circles, respectively ({\bf c});
Scd, Sd and irregular/BCD galaxies with empty pentagons, stars and crosses,
respectively, and generic peculiar and irregular galaxies with broad triangles
({\bf d}).
In each panel, we represent characteristic observational errors with crosses
and we reproduce the Fundamental Plane relation (continuous line in the
$\kappa_1$--$\kappa_3$ plane), the borderline of the ``Zone of Exclusion''
(short-dashed line in the $\kappa_1$--$\kappa_2$ plane) and the distribution
of the E$+$S0 galaxies, within $\rm \pm 1 \sigma$ from the mean,
in the $\kappa_2$--$\kappa_3$ plane.
For E and S0 galaxies, we represent also the expected increase
of the three $\kappa$-coordinates when the kinetic energy of these stellar
systems is not negligible.
Galaxies of later Hubble type march farther away from
the ``Zone of Exclusion'', while spiral and irregular/BCD galaxies
have higher mean effective dynamical mass-to-near-IR light ratios
than E$+$S0$+$S0a galaxies.
\newline
{\bf Fig. 3:} The galaxy distribution in the $r_e$--$\mu_e$ plane,
according to the decomposition of their radial light profiles or morphology.
Here, we represent: pure de Vaucouleurs systems with filled hexagons
({\bf a}); E and S0 galaxies with filled circles and empty squares,
respectively ({\bf b}); pure exponential disk-systems and truncated-disk
systems with empty triangles ({\bf c}); de Vaucouleurs/exponential
bulge$+$disk systems with filled squares ({\bf d}).
In each panel, the short-dashed line reproduces the Kormendy relation
(Equ. 15) for the de Vaucouleurs systems.
E$+$S0 galaxies follow this relation, whatever their profile decompositions
are.
Conversely, pure exponential disk-systems and truncated-disk systems
do not populate the region of the $r_e$--$\mu_e$ plane beyond this line.
Bulge$+$disk systems populate the region of the $r_e$--$\mu_e$ plane
immediately below this line.
\vskip 2.0truecm
\bsp

\label{lastpage}


\begin{thebibliography}{95}
\bibitem{b1} Aaronson M., Huchra J., Mould J., 1979, ApJ, 229, 1
\bibitem{b2} Avila--Reese V., Firmani C., 2000, in ``The seventh Texas--Mexico
conference on astrophysics: flows, blows, and glows'', (Eds.: W.H. Lee
and S. Torres--Peimbert), Rev. Mex. Astron. Astrofis. Serie de Conf. vol. 10,
p.97
\bibitem{b3} Baes M., Dejonghe H., 2000, MNRAS, 313, 153
\bibitem{b4} Baes M., Dejonghe H., De Rijcke S., 2000, MNRAS, 318, 798
\bibitem{b5} Bender R., Burstein D., Faber S.M., 1992, ApJ, 399, 462
\bibitem{b6} Bianchi S., Ferrara A., Davies J.I., Alton P.B., 2000, MNRAS,
311, 601
\bibitem{b7} Binggeli B., Sandage A., Tammann G.A., 1985, AJ, 90, 1681 (VCC)
\bibitem{b8} Binney J., Tremaine S., 1987, ``Galactic dynamics'',
Princeton University Press
\bibitem{b9} Bertola F., Galletta G., 1978, ApJ, 226, L115
\bibitem{b10} Boselli A., Gavazzi G., Donas J., Scodeggio M., 2001,
AJ, 121, 753
\bibitem{b11} Boselli A., Gavazzi G., Franzetti P., Pierini D.,
Scodeggio M., 2000a, A\&AS, 142, 73 (Paper IV)
\bibitem{b12} Boselli A., Tuffs R.J., Gavazzi G., Hippelein H.,
Pierini D., 1997, A\&AS, 121, 507 (B97)
\bibitem{b13} Burstein D., 1979, ApJ, 234, 435 
\bibitem{b14} Burstein D., Bender R., Faber S.M., Nolthenius R., 1995,
Astron. Lett. and Commun., 31, 95
\bibitem{b15} Burstein D., Bender R., Faber S.M.,
Nolthenius R., 1997, AJ, 114, 1365 (BBFN)
\bibitem{b16} Busarello G., Capaccioli M., Capozziello S., Longo G.,
Puddu E., 1997, A\&A, 320, 415
\bibitem{b17} Caon N., Capaccioli M., D'Onofrio M., 1993, MNRAS, 265, 1013
\bibitem{b18} Capelato H.V., de Carvalho R.R., Carlberg R.G., 1995, ApJ,
451, 525
\bibitem{b19} Capelato H.V., de Carvalho R.R., Carlberg R.G., 1997,
in ``Galaxy Scaling Relations: Origins, Evolution and Applications'',
(Eds.: L.N. da Costa and A. Renzini), Springer-Verlag, p.33
\bibitem{b20} Cardelli J.A., Clayton G.C., Mathis J.S., 1989, ApJ, 345, 245
\bibitem{b21} Dalcanton J.J., Spergel D.N., Summers F.J., 1997, ApJ, 482, 659
\bibitem{b22} de Jong R.S., 1996a, A\&A, 313, 45
\bibitem{b23} de Jong R.S., 1996b, A\&A, 313, 377
\bibitem{b24} de Vaucouleurs G., 1948, Ann. d'Astrophys., 11, 247
\bibitem{b25} de Vaucouleurs G., 1959, Hand. der Physik, 53, 311
\bibitem{b26} de Vaucouleurs G., 1977, ApJS, 33, 211
\bibitem{b27} de Zeeuw T., Franx M., 1991, ARA\&A, 29, 239
\bibitem{b28} Disney M.J., Davies J.I., Phillipps S., 1989, MNRS, 239, 939 
\bibitem{b29} Djorgovski S., Davis M., 1987, ApJ, 313, 59
\bibitem{b30} Dressler A., Lynden-Bell D., Burstein D., Davies R.L.,
Faber S.M., Terlevich R., Wegner G., 1987, ApJ, 313, 42
\bibitem{b31} Faber S.M., Jackson R.E., 1976, ApJ, 204, 668
\bibitem{b32} Ferrara A., Bianchi S., Cimatti A., Giovanardi C., 1999,
ApJS, 123, 437 
\bibitem{b33} Firmani C., Avila--Reese V., 2000, MNRAS, 315, 457
\bibitem{b34} Freeman K.C., 1970, ApJ, 160, 811
\bibitem{b35} Gavazzi G., Boselli A., 1996, Astroph. Lett. \& Comm., 35, 1 
\bibitem{b36} Gavazzi G., Scodeggio M., 1996, A\&A, 312, L29
\bibitem{b37} Gavazzi G., Pierini D., Boselli A., 1996a, A\&A, 312, 397 (GPB)
\bibitem{b38} Gavazzi G., Pierini D., Boselli A., Tuffs R.J.,
1996b, A\&AS, 120, 489 (Paper I)
\bibitem{b39} Gavazzi G., Pierini D., Baffa C., Lisi F., Hunt L.K.,
Randone I., Boselli A., 1996c, A\&AS, 120, 521 (Paper II)
\bibitem{b40} Gavazzi G., Boselli A., Scodeggio M., Pierini D.,
Belsole E., 1999, MNRAS, 304, 595
\bibitem{b41} Gavazzi G., Franzetti P., Scodeggio M., Boselli A.,
Pierini D., 2000b, A\&A, 361, 863 (Paper V)
\bibitem{b42} Gavazzi G., Franzetti P., Scodeggio M., Boselli A., Pierini D.,
Baffa C., Lisi F., Hunt L.K., 2000a, A\&AS, 142, 65 (Paper III)
\bibitem{b43} Gavazzi G., Zibetti S., Boselli A., Franzetti P., Scodeggio M.,
Martocchi S., 2001, A\&A, 372, 29 (Paper VI)
\bibitem{b44} Geha M., Guhathakurta P., van der Marel R., 2001,
astro-ph/0107010
\bibitem{b45} Graham A., Colless M., 1997, MNRAS, 287, 221
\bibitem{b46} Grosb$\rm \o$l P.J., 1985, A\&AS, 60, 261
\bibitem{b47} Hawarden T.G., Longmore A.J., Tritton S.B., Elson R.A.W.,
Corwin H.G., 1981, MNRAS, 196, 747
\bibitem{b48} Haynes M.P., Giovanelli R., 1984, AJ, 89, 758
\bibitem{b49} Hubble E., 1926, ApJ, 64, 321
\bibitem{b50} Jura M., 1986, ApJ, 306, 483
\bibitem{b51} Kennicutt R.C., Tamblyn P., Congdon C.W., 1994, ApJ, 435, 22
\bibitem{b52} Khosroshahi H.G., Wadadekar Y., Kembhavi A., 2000,
ApJ, 533, 162 
\bibitem{b53} Khosroshahi H.G., Wadadekar Y., Kembhavi A., Mobasher B.,
2000, ApJ, 531, L103
\bibitem{b54} King I.R., 1966, AJ, 71, 64
\bibitem{b55} Kormendy J., 1977, ApJ, 218, 333
\bibitem{b56} Kormendy J., 1988, in ``Origin, Structure and Evolution
of Galaxies'', IAU Symposium 127, (Ed.: L.Z. Fang), World Scientific,
Singapore, p.252
\bibitem{b57} Lequeux J., 1983, A\&A, 125, 394 
\bibitem{b58} Mao S., Mo H.J, 1998, MNRAS, 296, 847
\bibitem{b59} Mao S., Mo H.J, White S.D.M., 1998, MNRAS, 297, L71
\bibitem{b60} Maraston C., 1998, MNRAS, 300, 872
\bibitem{b61} Mo H.J, Mao S., White S.D.M., 1998, MNRAS, 295, 319
\bibitem{b62} Mobasher B., Guzman R., Aragon--Salamanca A., Zepf S., 1999,
MNRAS, 304, 225
\bibitem{b63} Moriondo G., Giovanelli R., Haynes M.P., 1999, A\&A, 346, 415
\bibitem{b64} Moriondo G., Baffa C., Casertano S., Chincarini G., Gavazzi G.,
Giovanardi C., Hunt L.K., Pierini D., Sperandio M., Trinchieri G., 
1999, A\&AS, 137, 101
\bibitem{b65} Pahre M.A., 1999, ApJS, 124, 127
\bibitem{b66} Pahre M.A., de Carvalho R.R., Djorgovski S.G., 1998, AJ, 116,
1606
\bibitem{b67} Pahre M.A., Djorgovski S.G., de Carvalho R.R.,
1998, AJ, 116, 1591 (PDdC)
\bibitem{b68} Pahre M.A., Djorgovski S.G., de Carvalho R.R., 1999,
in ``Star Formation in Early Type Galaxies'', ASP Conference Series 163,
(Eds.: P. Carral and J. Cepa), p. 17
\bibitem{b69} Peletier R.F., Balcells M., Davies R.L., Andredakis Y.,
Vazdekis A., Burkert A., Prada F., 1999, MNRAS, 310, 703
\bibitem{b70} Persic M., Salucci P., Stel F., 1996, MNRAS, 281, 27
\bibitem{b71} Pierini D., 1997, Ph.D. thesis, Univ. of Milano
\bibitem{b72} Pierini D., Tuffs R.J., 1999, A\&A, 343, 751
\bibitem{b73} Renzini A., 1998, in ``The Young Universe: Galaxy Formation
and Evolution at Intermediate and High Redshift'', (Eds.: S. D'Odorico,
A. Fontana and E. Giallongo), ASP Conference Series 146, p.298
\bibitem{b74} Richter O.-G., Huchtmeier W.K., 1984, A\&A, 132, 253
\bibitem{b75} Rix H.-W., Carollo M., Freeman K., 1999, ApJ, 513, L25
\bibitem{b76} Roberts M., Haynes M., 1994, ARA\&A, 32, 115
\bibitem{b77} Rubin V.C., Burstein D., Ford W.K., Thonnard N.,
1985, ApJ, 289, 91
\bibitem{b78} Saglia R.P., Bender R., Dressler A., 1993, A\&A, 279, 75
\bibitem{b79} Sandage A., 1961, ``The Hubble Atlas of Galaxies'',
Carnagie Institute of Washington DC
\bibitem{b80} Sandage A., 1986, A\&A, 161, 89
\bibitem{b81} Scodeggio M., 2001, AJ, 121, 2413
\bibitem{b82} Scodeggio M., Gavazzi G., 1993, ApJ, 409, 110
\bibitem{b83} Scodeggio M., Gavazzi G., Belsole E., Pierini D., Boselli A.,
1998, MNRAS, 301, 1001 (S98)
\bibitem{b84} Scodeggio M., Gavazzi G., Franzetti P., Zibetti S., Boselli A.,
Pierini D., 2001, A\&A, in press (Paper IX)
\bibitem{b85} Silk J., 1997, ApJ, 481, 703
\bibitem{b86} Simien F., Prugniel P., 2000, A\&AS, 145, 263
\bibitem{b87} Syer D., Mao S., Mo H.J., 1999, MNRAS, 305, 357
\bibitem{b88} Tammann G.A., 1980, in ``ESO/ESA Workshop on Dwarf Galaxies'',
Knudsen, (Eds.: M. Tarenghi and K. Khar), p. 3
\bibitem{b89} Teerikorpi P., 1987, A\&A, 173, 39 
\bibitem{b90} Teerikorpi P., 1990, A\&A, 234, 1
\bibitem{b91} Tully R.B., Fisher J.R., 1977, A\&A, 54, 661
\bibitem{b92} Whitmore B.C., Kirshner R.P., 1981, ApJ, 250, 43
\bibitem{b93} Witt A.N., Thronson H.A., Capuano J.M., 1992, ApJ, 393, 611
\bibitem{b94} Zaritsky D., Kennicutt R., Huchra J., 1994, ApJ, 420, 87
ASP Conference Series 8, p.1
\bibitem{b95} Zwicky F., Herzog E., Karpowicz M., Kowal C., Wild P.,
1961--1968, ``Catalogue of Galaxies and Clusters of Galaxies'', vol. 6,
Pasadena, C.I.T. (CGCG)
\end{thebibliography}
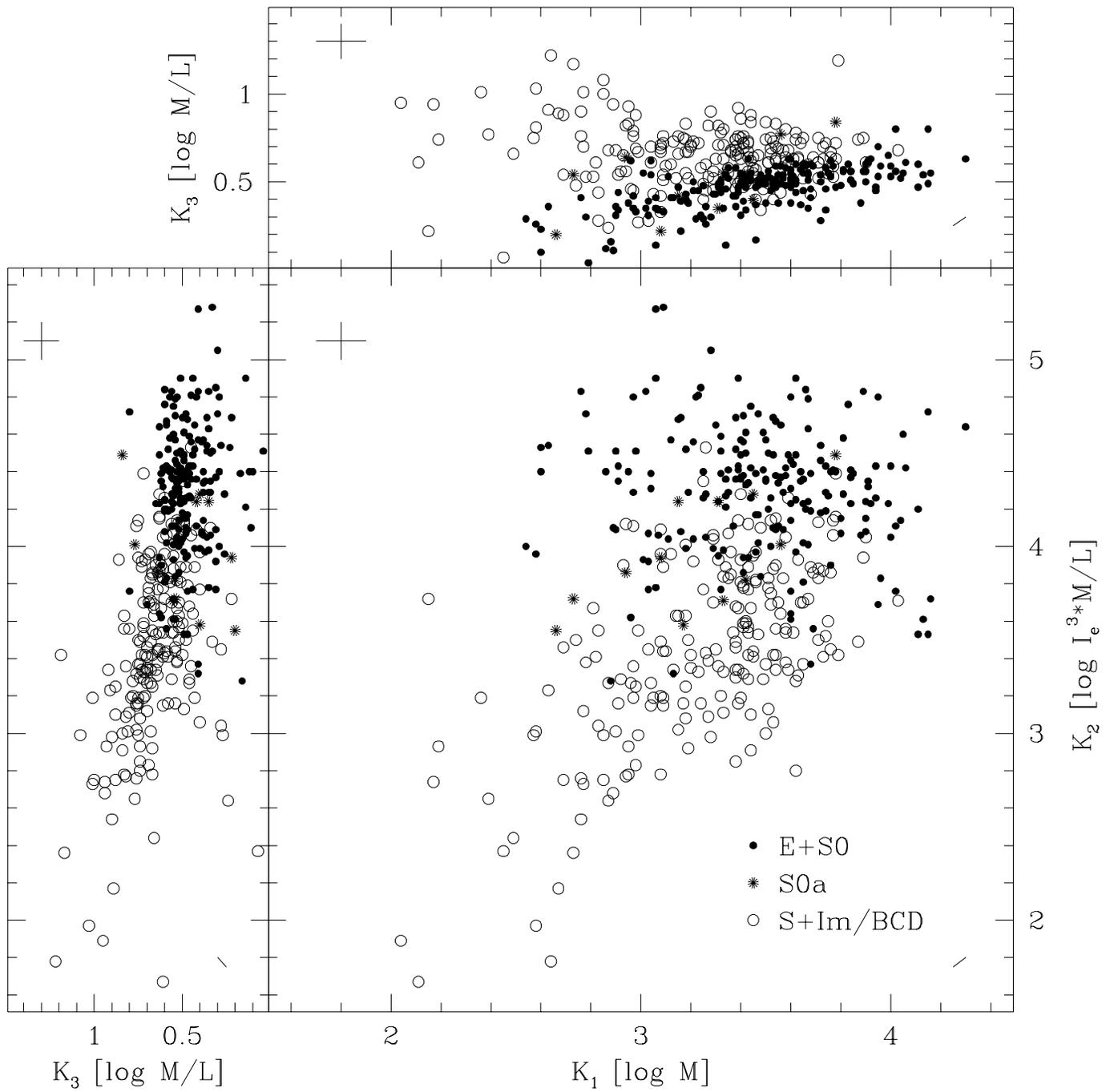
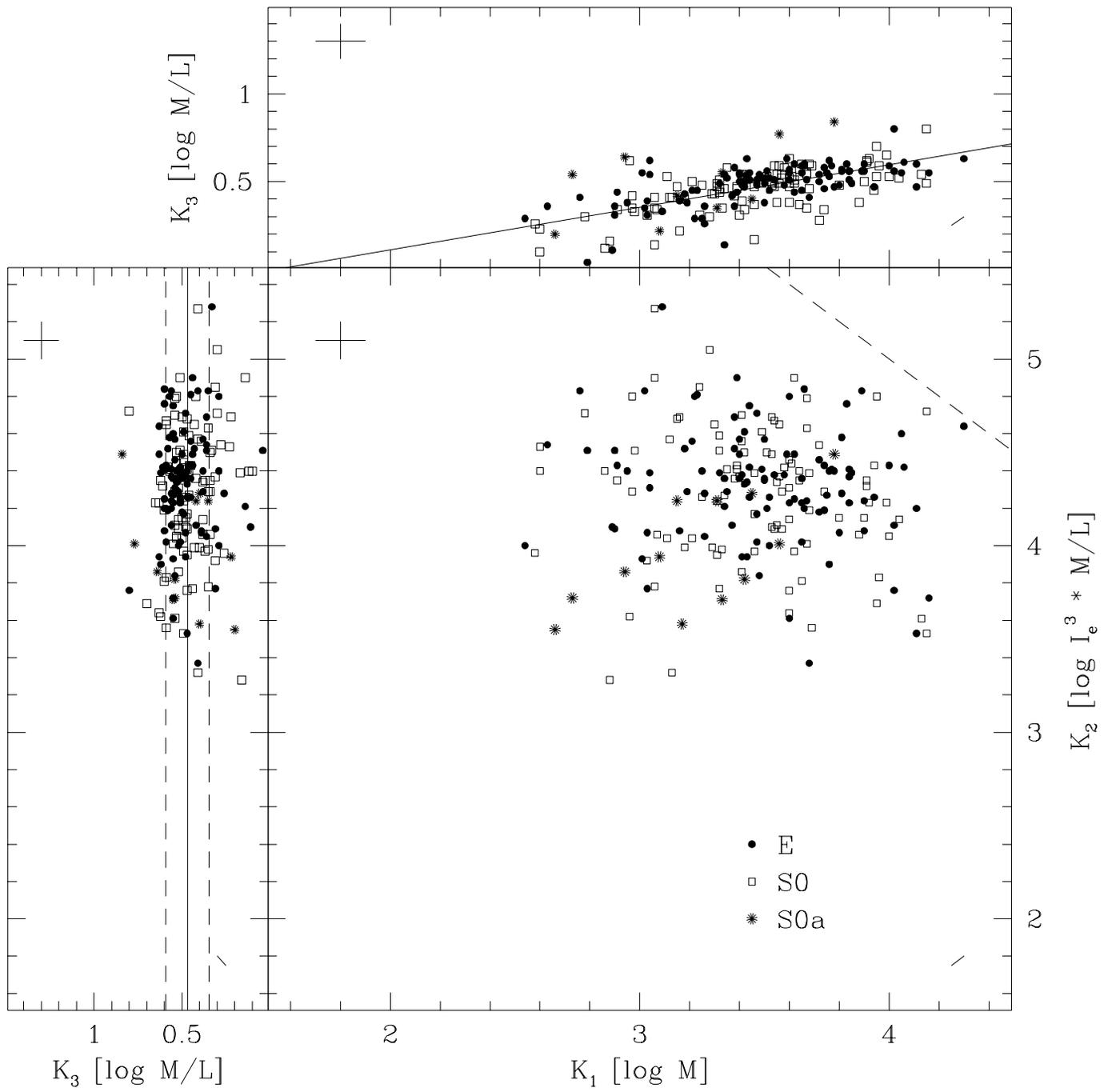
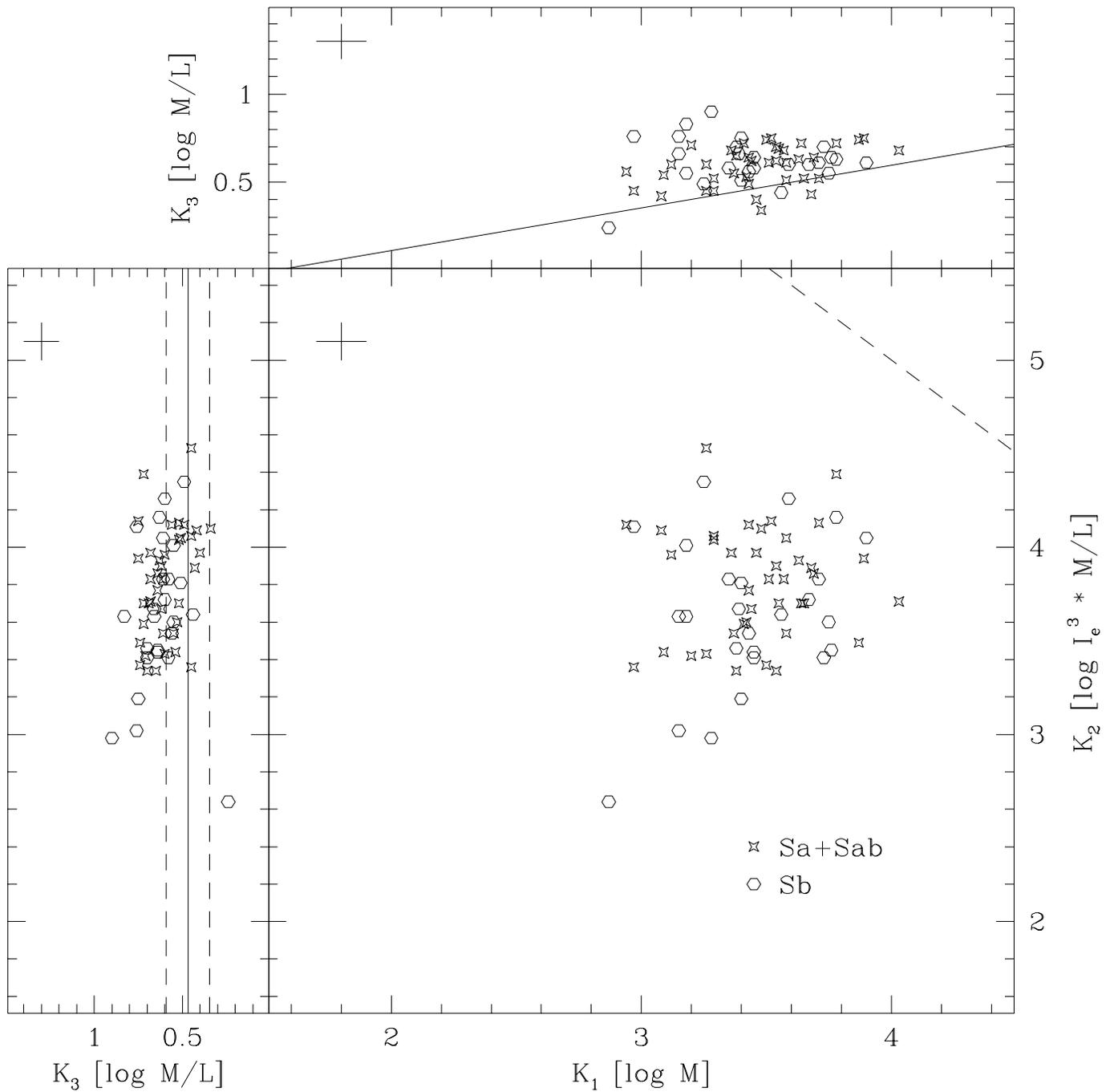
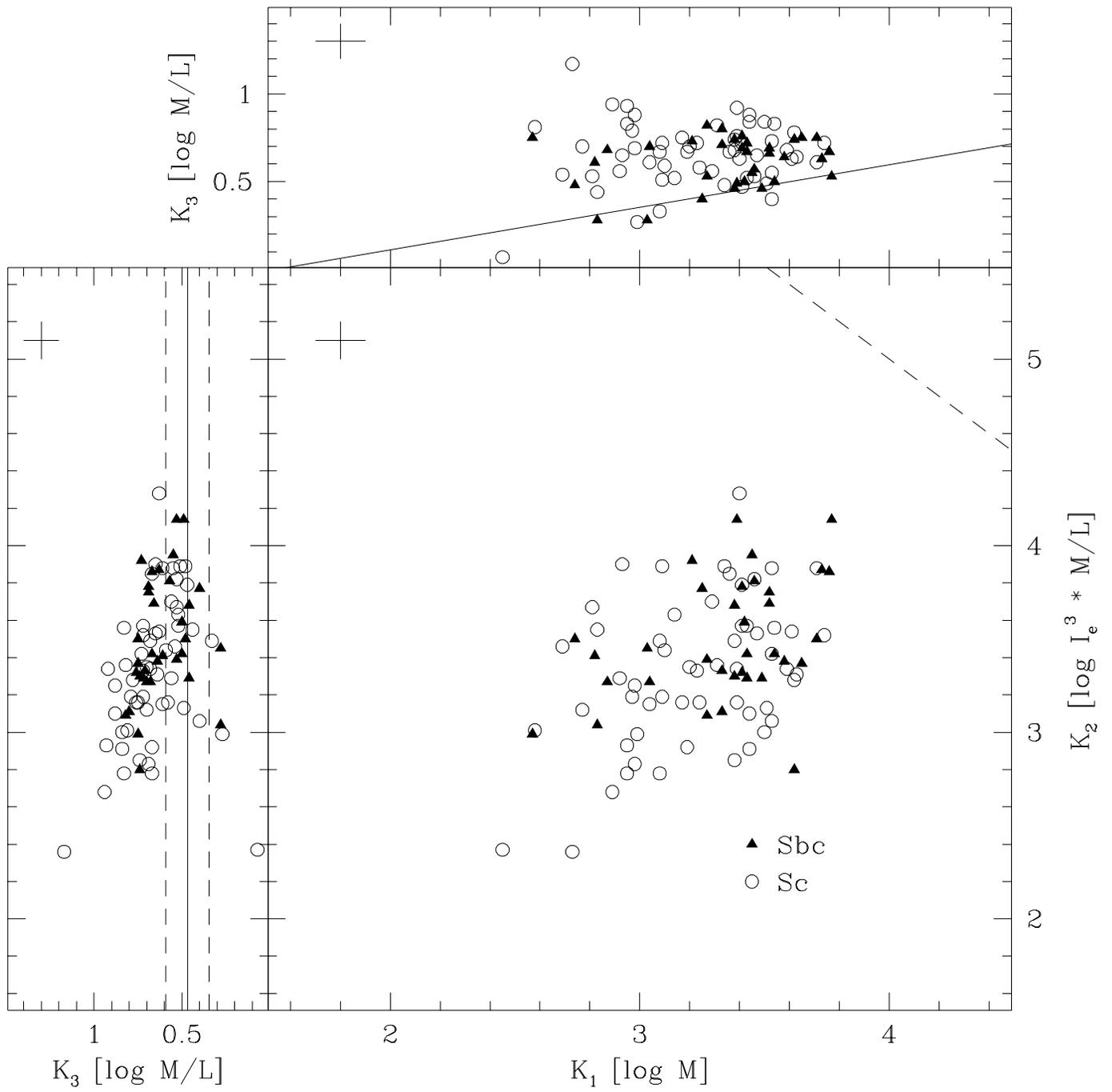
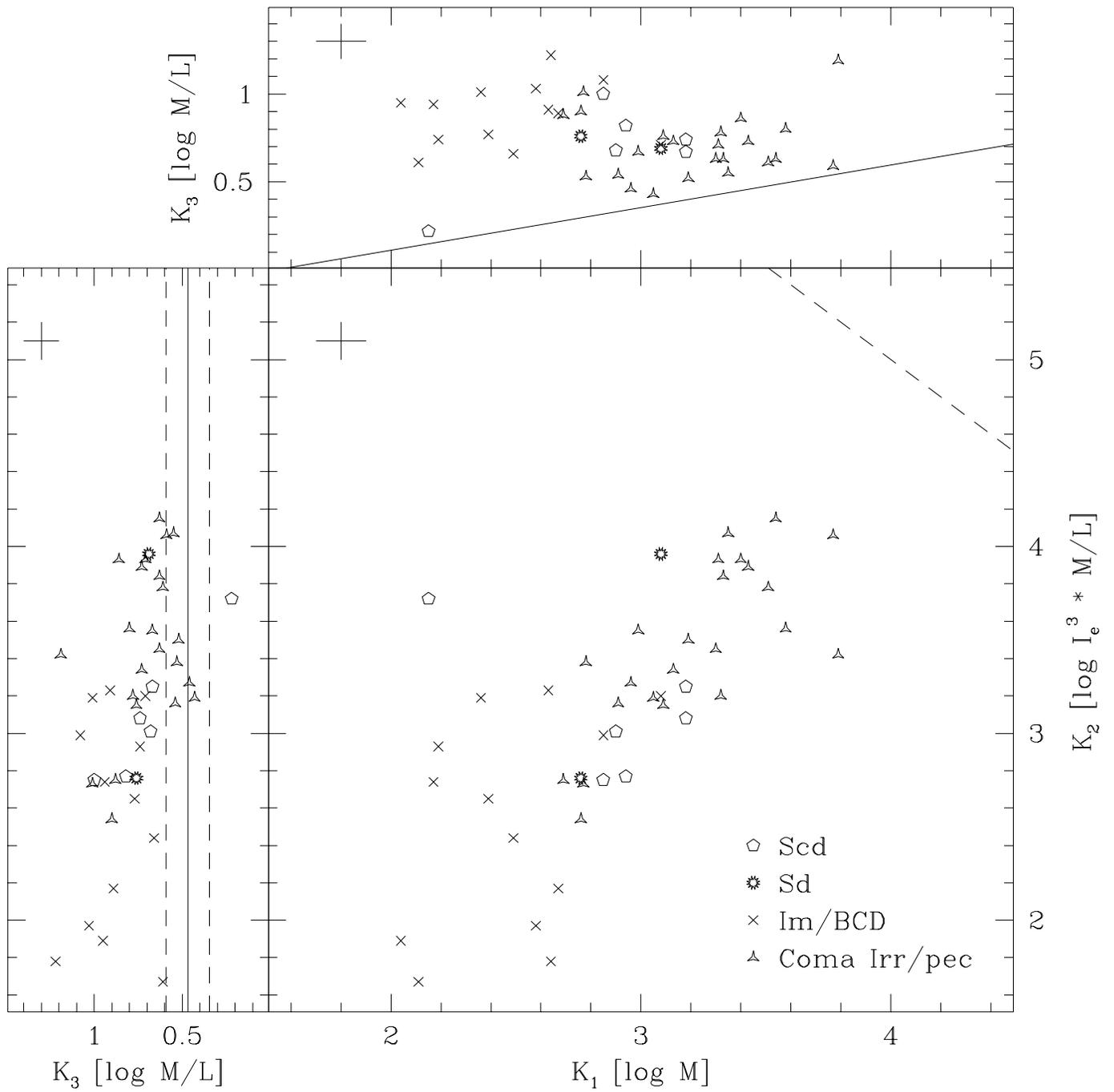
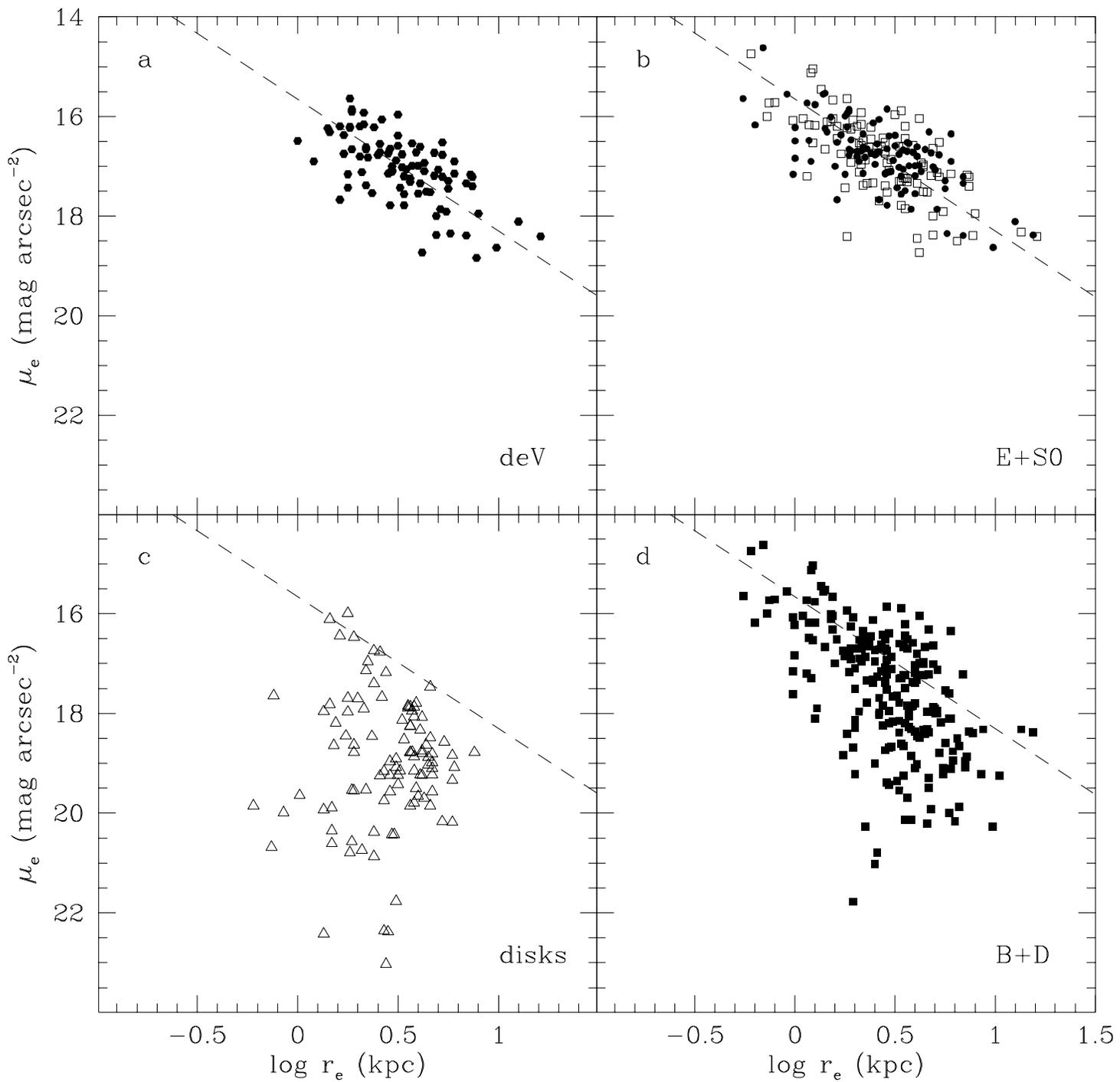
\end{document}